\documentclass[doublecol]{epl2}

\title{Charge creation and nucleation of longitudinal plasma wave in coupled Josephson junctions}
\shorttitle{Title} 

\author{Yu. M. Shukrinov\inst{1,2} \and M.Hamdipour\inst{3}}
\shortauthor{Yu. M. Shukrinov \etal}

\institute{
  \inst{1} BLTP, JINR, Dubna, Moscow Region, 141980, Russia\\
  \inst{2} Max-Planck-Institute for the Physics of Complex Systems, 01187 Dresden, Germany\\
  \inst{3} Institute for Advanced Studies in Basic Sciences, P.O.Box 45195-1159, Zanjan, Iran
} \pacs{74.50.+r}{Tunneling phenomena; Josephson effects} \pacs{74.81.Fa}{Josephson junction arrays and wire
networks}

\abstract{ We study the phase dynamics in coupled Josephson junctions describing by system of nonlinear
differential equations. Results of detailed numerical simulations of charge creation in the superconducting
layers and the longitudinal plasma wave (LPW) nucleation  are presented. We demonstrate the different time
stages in the development of the LPW and present results of FFT analysis at different values of bias current.
The correspondence between the breakpoint position on the outermost branch of current voltage characteristics
(CVC) and the growing region in time dependence of the electric charge in the superconducting layer is
established. The effects of noise in the bias current and the external microwave radiation on the charge
dynamics of the coupled Josephson junctions are found. These effects introduce a way to regulate the process of
LPW nucleation in the stack of IJJ.}

\begin{document}

\maketitle

\section{Introduction}

The intrinsic Josephson junctions (IJJ) in high temperature superconductors (HTSC)  are formed from the atomic
scale superconducting layers (S-layers), with each layer having a thickness that is comparable with the Debye
screening length.\cite{kleiner92} Because of it, the electric charge does not screen  in S-layers perfectly and
it leads to the capacitive coupling between junctions and the existence of the LPW along the
stack.\cite{koyama96, machida98,sm-prl07}

The inductive coupling\cite{sakai93} of IJJ in the absence of magnetic field can be neglected for the small size
stacks,  and the phase dynamics in such nanojunctions is determined by the capacitive coupling only. In this
case the system of IJJ is described by capacitively coupled Josephson junctions (CCJJ) model\cite{koyama96} or
the model with the diffusion current (CCJJ+DC)\cite{machida00,sms-phC06}. The investigation of the charge
dynamics in such stack  allowed us to predict new physical properties of the coupled system of
JJ.\cite{sm-prl07,sms-prb08}  Its CVC is characterized by the multiple branch structure  and branches have a
breakpoint and some breakpoint region (BPR) before the transition to another branch.\cite{sm-prl07}

The physical properties of IJJ are investigating very intensively today. The collective Josephson plasma
resonance was observed experimentally some time ago in Ref.\cite{matsuda95} but the observation of the powerful
coherent radiation from the stack of IJJ\cite{ozyuzer} and the experimental manifestation of the breakpoint and
the BPR\cite{irie} in CVC of  $Bi_2Sr_2CaCu_2O_y$ stimulates new investigations in this field. The key problems
are the understanding of the mechanism of this radiation and the way to increase its power. An important
question concerns the nucleation of the LPW in the stack of IJJ which is not investigated yet. The
correspondence between the breakpoint's position in CVC and the parametric resonance region in time dependence
of the electric charge in the S-layers is not precisely established as well. The important open issue is a
question if it is possible to regulate the process of LPW nucleation.

In this paper we investigate the phase dynamics in the coupled system of Josephson junctions. We show the
different time stages of charge creation in S-layers and the development of the LPW. The data concern the charge
distribution along the stack and the results of fast Fourier transformation analysis at different values of bias
current.  It is found that the nucleation process of LPW can be affected by the noise in bias current  and the
external microwave radiation. It makes possible to regulate the process of LPW nucleation  in HTSC.

\section{CVC and time dependence of the charge in superconducting layer}

To simulate the CVC of IJJ, we solve a system of dynamical equations $d^2\varphi_{l}/d\tau^2=(I-\sin \varphi_{l}
-\beta d\varphi_{l}/d\tau)+ \alpha (\sin \varphi_{l+1}+ \sin\varphi_{l-1} - 2\sin\varphi_{l})+ \alpha \beta(
d\varphi_{l+1}/d\tau+d\varphi_{l-1}/d\tau-2d\varphi_{l}/d\tau)$ for the gauge-invariant phase differences
$\varphi_l(\tau)= \theta_{l+1}(\tau)-\theta_{l}(\tau)-\frac{2e}{\hbar}\int^{l+1}_{l}dz A_{z}(z,\tau)$ between
superconducting layers ($S$-layers), where $\theta_{l}$ is the phase of the order parameter in the S-layer $l$,
$A_z$ is the vector potential in the barrier and $\alpha$ is the coupling parameter. The third  term in the
right hand side is the diffusion current which is determined by the difference of the generalized scalar
potentials of superconducting layers.\cite{koyama96, ryndyk98} In our simulations we measure the voltage in
units of $V_0=\hbar\omega_p/(2e)$ and the current in units of the $I_c$. We use a dimensionless time $\tau =
t\omega_p$, where $\omega_{p}$ is the plasma frequency $\omega_{p}=\sqrt{2eI_c/\hbar C}$, ${I_c}$ is the
critical current and $C$ is the capacitance. The CVC and time dependence of the charge oscillation in the
S-layers are simulated at coupling parameter $\alpha = 1$, dissipation parameter $\beta = 0.2$ and periodic BC.
We chose arbitrary initial value of bias current to record the time dependence. The time dependence consists of
time and bias current variations. At each current step the phase dynamics is developed in the definite time
interval. A small noise  with its maximum in the interval $(-10^{-8}, +10^{-8})$ is added in bias current in our
simulations. The details concerning the model and numerical procedure are presented in
Refs.\cite{sms-phC06,sm-prl07}

\begin{figure}[ht!]
 \centering
\includegraphics[height=60mm]{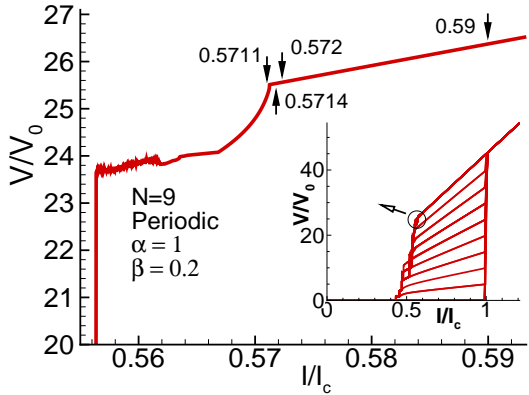}
\caption{(Color online) The enlarged part of the outermost branch of the CVC of the stack wit 9 IJJ. The inset
shows the total CVC.}\label{1}
\end{figure}
To compare our results with previous investigations of the CVC and charge dynamics in the coupled system of JJ
we concentrate here on the stack with 9 JJ. A correspondence between the CVC and charge oscillations  in the
S-layers in such stack was demonstrated in Ref.\cite{sms-prb08} It was shown that at chosen values of $\alpha$
and $\beta$ the LPW with wave vector $k=8\pi/9$ is created in the stacks with 9 coupled JJ at the
breakpoint.\cite{sm-prl07} Here we study the nucleation process of LPW which was not touched before.

The CVC of the stack with 9 coupled JJ is presented in the inset to Fig.~\ref{1}. It has a large hysteretic
region with 9 branches. We show by circle the BPR on the outermost branch and its enlarged part in this figure.
The arrows indicate the bias current values at which the detailed study of the charged dynamics have been done.
We investigate the dependence of the electric charge in S-layer vs time at the following values of bias current
$I=0.59, 0.572, 0.5714$ and $I=0.5711$. We have done the FFT analysis of the charges oscillations at these
values of current as well. As we mention in Ref.\cite{sms-prb08}, the positions of the breakpoint in CVC and the
onset of charge growth in S-layer in time dependence are not coincide.

\section{Time stages in the nucleation of LPW}
A correspondence between the breakpoint in CVC and charge dynamics in S-layers was demonstrated in the previous
research.\cite{sms-prb08} Here we demonstrate that the creation and development of LPW has different stages in
time (or current). We show that the breakpoint in CVC reflects just the region of sharp increase of the LPW
amplitude. Investigation of the time dependence of electric charge in the S-layers give us the information
concerning the nucleation of LPW.

At chosen values of $\alpha$ and $\beta$, the LPW with wave vector $k=8\pi/9$ is created at the
breakpoint.\cite{sm-prl07} Below we show the profiles of the charge oscillations and some enlarged parts for
different values of bias current indicated in Fig.~\ref{1}.

\subsection{Fluctuation region}
\begin{figure}[ht!]
 \centering
\includegraphics[height=30mm]{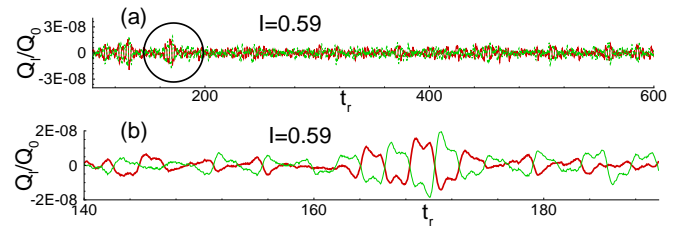}
\caption{(Color online) (a) The characteristic part of dependence the charge in first two S-layers vs time at
$I=0.59$ for the stack with 9 coupled JJ; (b) The enlarged part of the time interval (140,190) which
demonstrates the fluctuation region. The thick line shows the charge in the first layer of the stack, the thin
line corresponds to the charge in the second layer.}\label{2}
\end{figure}

Fig.~\ref{2}a shows the charge-time dependence for the first two S-layers of the stack with 9 coupled JJ at
$I=0.59$, which is far enough from the breakpoint in the outermost branch of CVC. We see that the charge value
$Q_l/Q_0$ is practically the same in both layers and it is about $10^{-8}$, i.e. it is on the noise level. The
charge oscillations here have mostly an irregular character. But nevertheless, as we can see in Fig.~\ref{2}b,
there are the regular oscillations in a short time interval with the frequency corresponding to the LPW. We call
this part in CVC and the corresponding part in time dependence of the charge on the S-layers as a fluctuation
region.

\begin{figure}[ht!]
 \centering
\includegraphics[height=65mm]{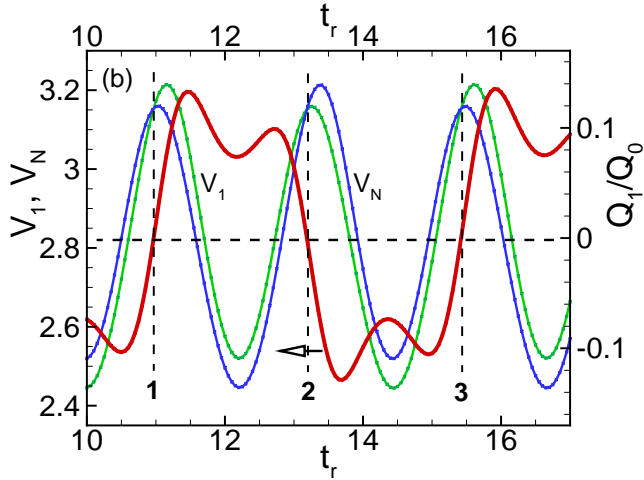}
\caption{(Color online) The  formation of the charge signal's shape in the first S-layer of the stack with nine
IJJ.}\label{3}
\end{figure}

Let us clarify the shape of the charge signal we see in Fig.~\ref{2}b . To do it, we demonstrate in Fig.~\ref{3}
the time dependence of the voltages in the first $V_1$ and last $V_N$ junctions (left $y$ axis), which are the
adjacent junctions to the first layer (we use the periodic BC and first layer has number 0). The charge density
$Q_l$ in the S-layer $l$ is expressed by the voltages $V_{l}$ and $V_{l-1}$ in the adjusting junctions $Q_l=Q_0
\alpha (V_{l}-V_{l-1})$, where $Q_0 = \varepsilon \varepsilon _0 V_0/r_D^2$, and $r_D$ is Debye screening
length. On the right $y$ axes we plot the value of charge in the first S-layer, which is actually equal to
$V_1-V_N$ at $\alpha=1$. At some time in during the oscillations the sign of charge in the first layer is
changed: the vertical dashed lines show the sign changes of the difference $V_{1}-V_{N}$.

We see that at time moment corresponding to the first dashed line,  the curve $V_1$ is getting above of curve
$V_N$, but at time moment corresponding to the second dashed line the situation is opposite and the charge in
the first layer is getting negative. The time dependence of the voltage difference, i.e. charge in the first
S-layer is shown by thick line. We see that the shape of the charge signal coincide qualitatively with the
charge signal observed in Fig.~\ref{2}b.  Of course, the shape of charge signal might have some deviations from
this one in the different layers. Its change in time reflects the character of Josephson junctions phase
dynamics at different values of bias current. So, by shifting of voltages along the time axis in the
Fig.~\ref{3}, we may reproduce  different shape of the charge signal ( particularly that one which appears in
Fig.~\ref{2}b).

\begin{figure}[ht!]
 \centering
\includegraphics[height=30mm]{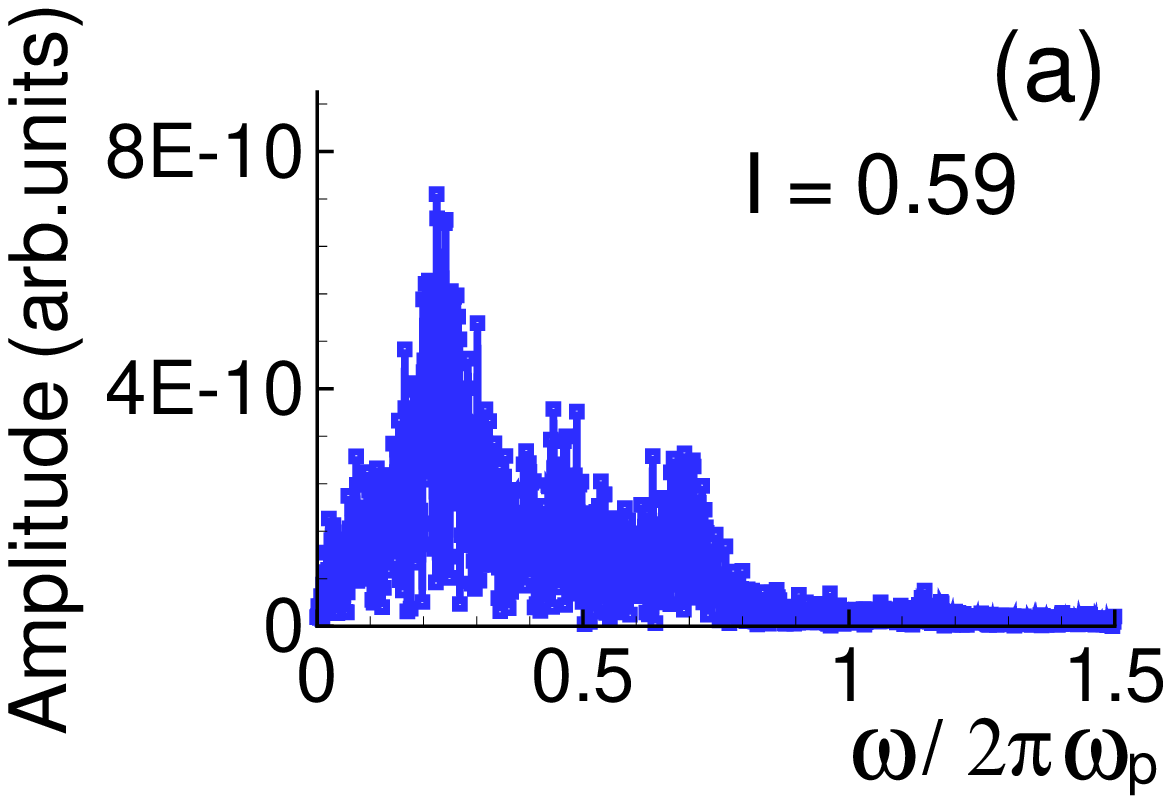}\includegraphics[height=35mm]{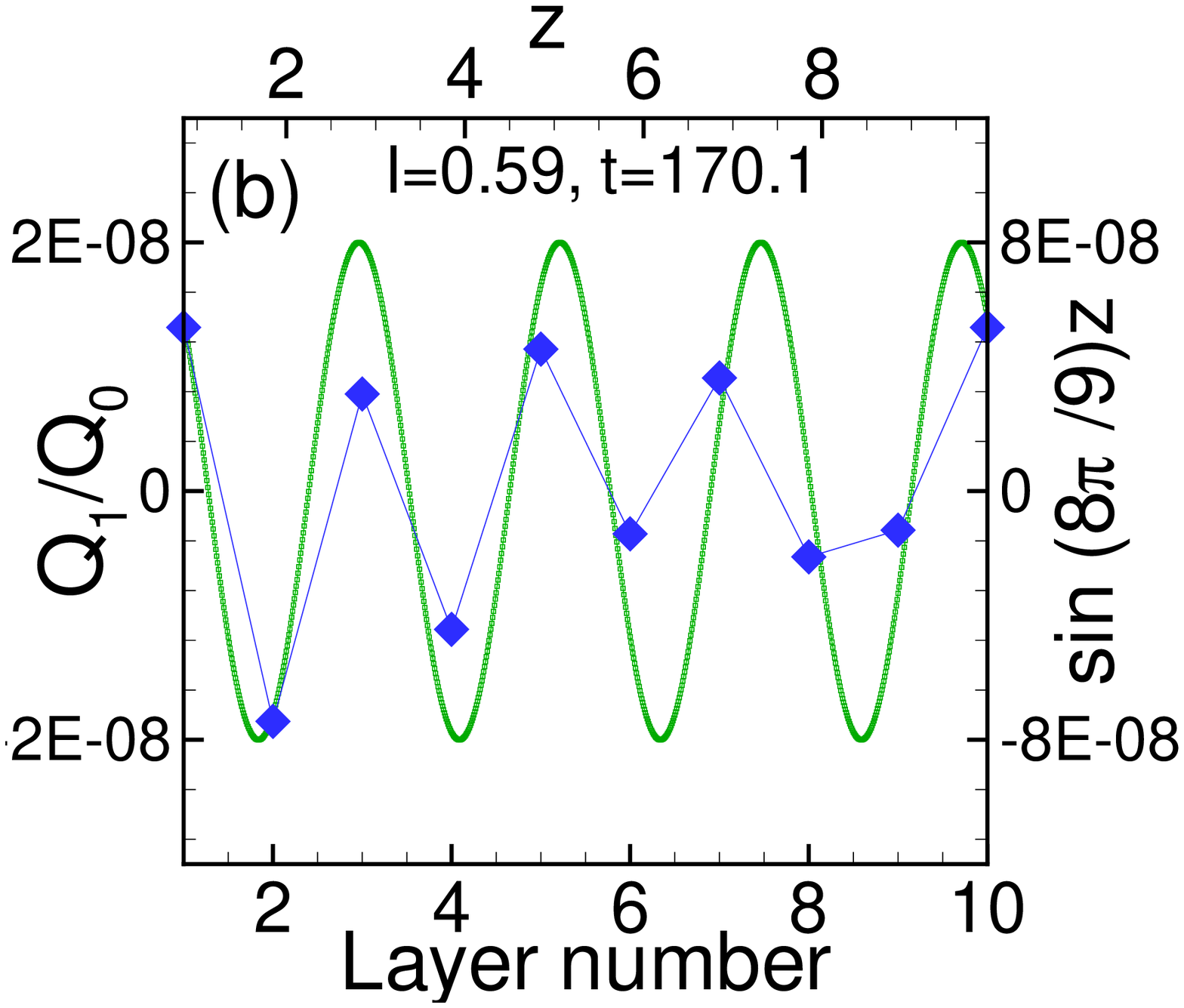}
\caption{(Color online) (a) Results of FFT analysis of the charge oscillations in the stack with nine IJJ at
bias current $I=0.59$; (b) Charge distribution along the stack at $I=0.59$.  The solid line in (b) shows the
wave $sin(8 \pi /9)$.}\label{4}
\end{figure}

The spectrum of Fast Fourier Transformation (FFT) analysis of the charge oscillations in the first S-layer in
the fluctuation region is shown in Fig.~\ref{4}a. As it's expected, it is broad and degraded. But nevertheless
it demonstrates the manifestation of the LPW (maximum in the interval 0.22-0.25) and the Josephson oscillations
(maximum in the interval 0.43-0.48).  Of course, the charge distribution along the stack deviates from the wave
with $k=8\pi/9$, as we can see in Fig.~\ref{4}b. The solid line shows the $\sin (8\pi/9)z$, the rhombus show the
charge value in the first S-layer $Q_1/Q_0$ at some fixed time moment $t=170.1$.

\subsection{Island region}

\begin{figure}[ht!]
 \centering
\includegraphics[height=30mm]{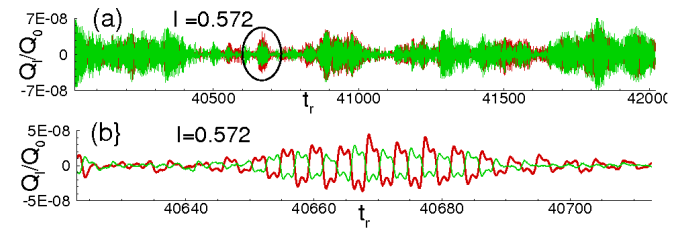}
\caption{(Color online) (a) The characteristic parts of charge time dependence in the stack with 9 IJJ in island
region; (b) The enlarged part of time dependence  marked by circle in figure (a). The thick line shows the
charge in the first S-layer of the stack, the thin line corresponds to the charge in the second S-layer.
}\label{5}
\end{figure}

With approaching the breakpoint on the outermost branch, the number of such intervals of regular oscillations we
demonstrated in Fig.~\ref{2}b is increasing and there size along time axes as well. The most important fact in
this case is the value of the oscillations amplitude. It  exceeds the noise value $10^{-8}$ essentially now. So,
we call such regions like the region shown in Fig.~\ref{5}a at $I=0.572$  as  the "island regions".

Fig.~\ref{5}a shows the charge-time dependence for first two S-layers of the stack in the island region and
Fig.~\ref{5}b shows the enlarged part marked by circle in this region. We see that the amplitude of charge
oscillations in shown island might have a value up to  $7*10^{-8}$.

\begin{figure}[ht!]
 \centering
\includegraphics[height=32mm]{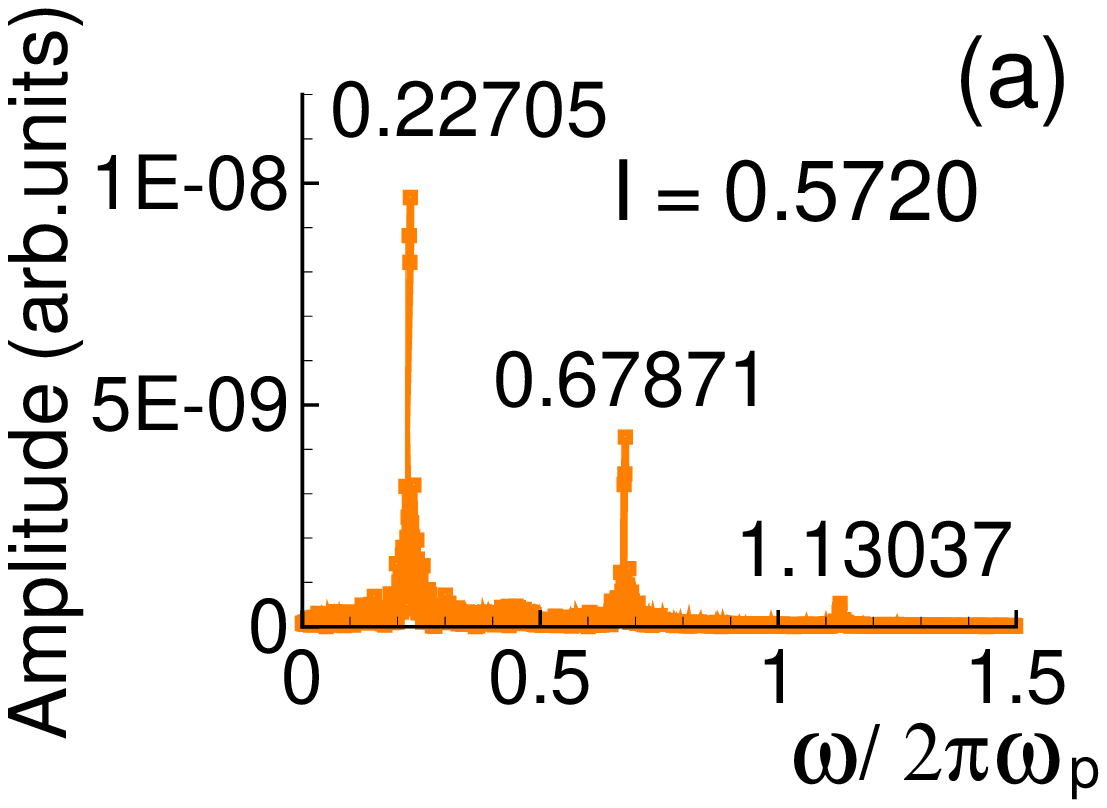}\includegraphics[height=32mm]{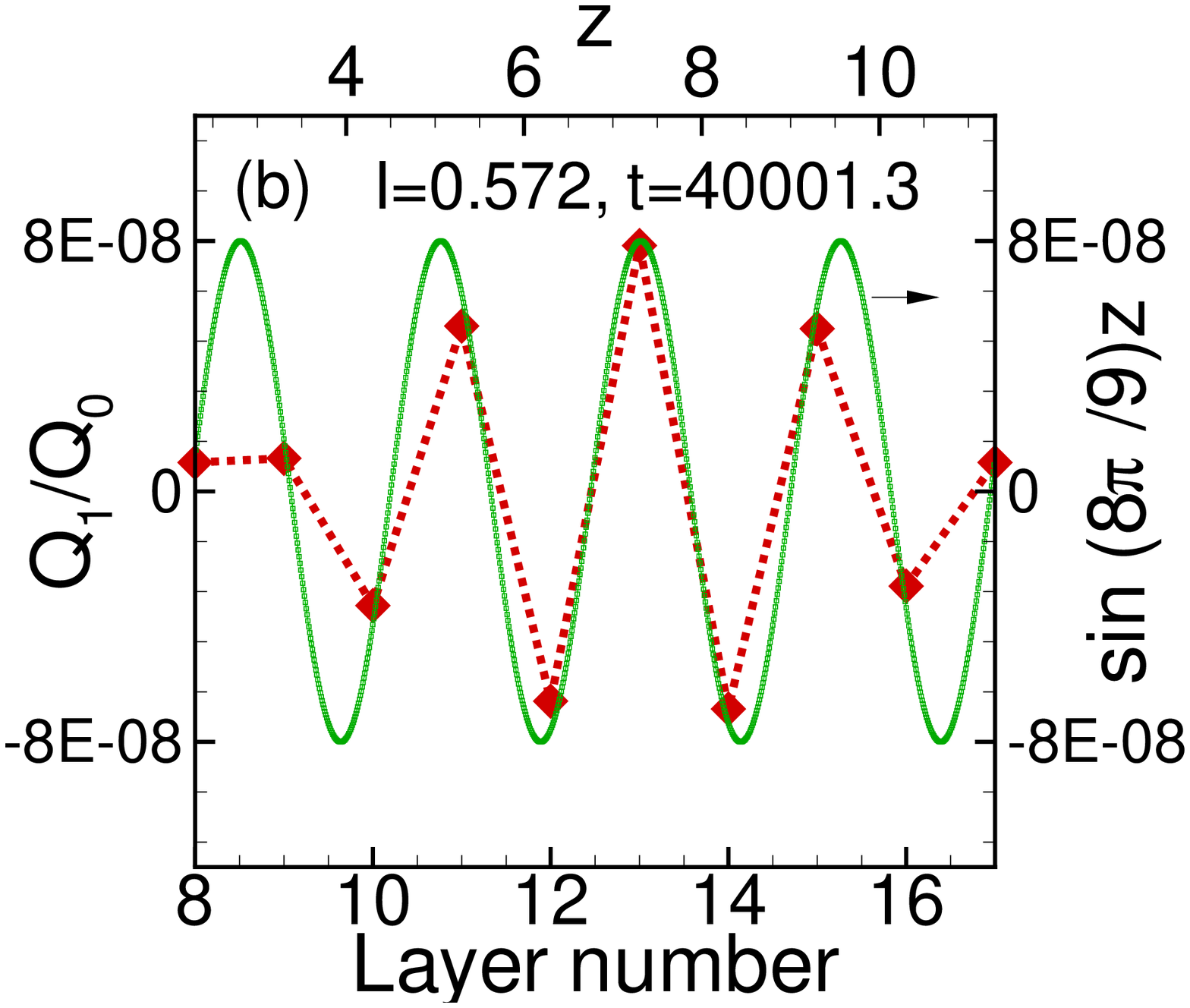}
\caption{(Color online) (a) Results of FFT analysis of the charge oscillations in the stack with nine coupled JJ
at bias current  $I=0.572$. (b) Charge distribution along the stack at  $I=0.572$. The solid line in (b) shows
the wave $sin(8 \pi /9)$.}\label{6}
\end{figure}

In the island region  we can precisely determine the LPW  frequency by the FFT analysis in contrast to the
fluctuation region. In Fig.~\ref{6}a we demonstrate the results of FFT analysis for charge oscillations at
$I=0.572$. As we can see, there is still the broadening part in the FFT spectrum, but the charge distribution
along the stack follows the modulated LPW with $k=8\pi/9$ (see Fig.~\ref{6}b ).

\subsection{Alternative amplitude region}
\begin{figure}[ht!]
 \centering
\includegraphics[height=25mm]{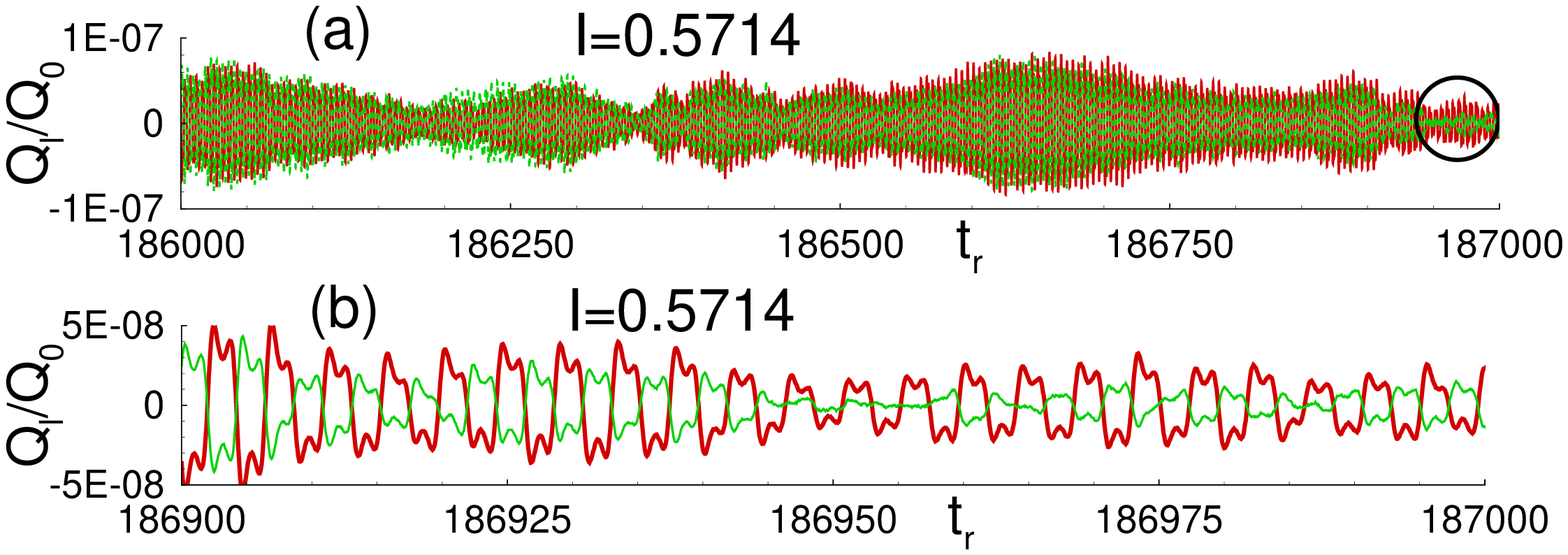}
\includegraphics[height=30mm]{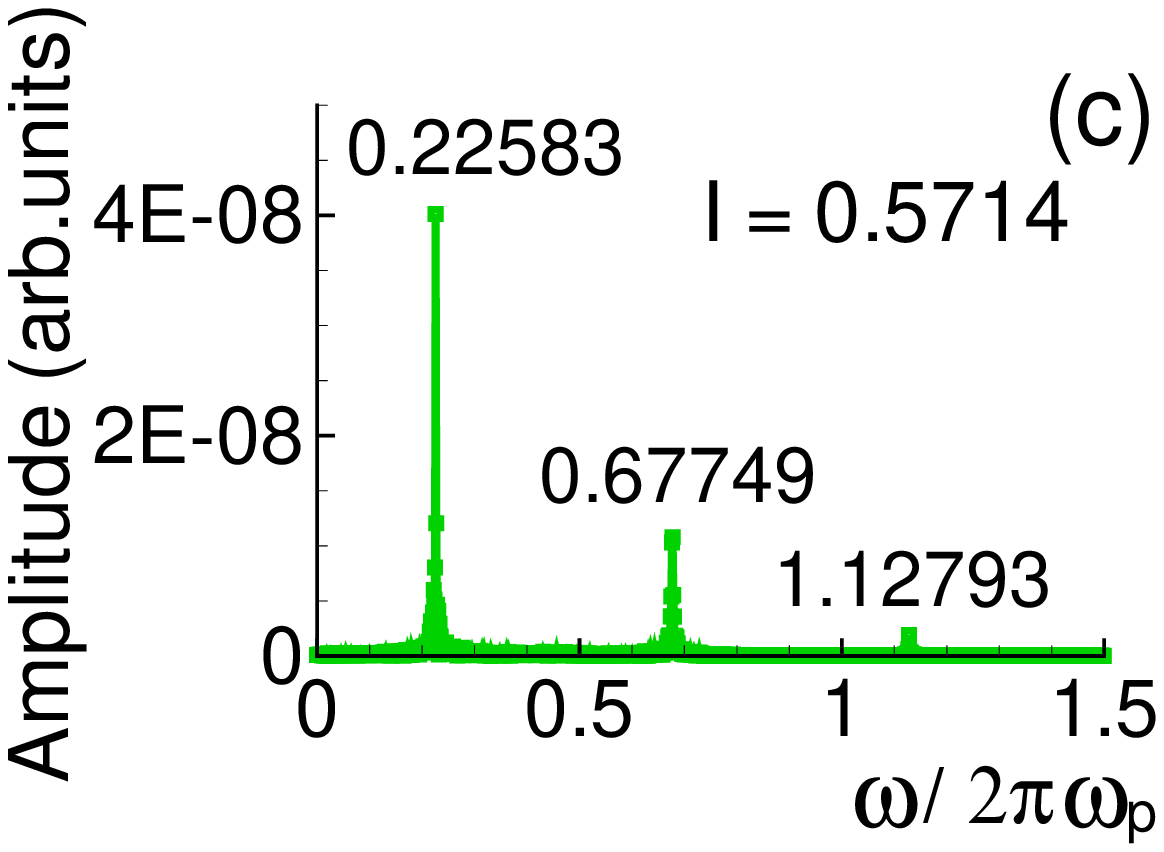}
\caption{(Color online) (a) The characteristic parts of charge-time dependence in the stack with 9 IJJ in the
alternative amplitude region; (b) Results of FFT analysis of the charge oscillations at bias current $I=0.5714$.
}\label{7}
\end{figure}

With increase in time (decrease in current) the islands are growing, joining, forming the region of charge
oscillations with alternating amplitude. The characteristic part of it is shown in Fig.~\ref{7}a. The amplitude
of oscillation here is around $10^{-7}$, i.e. one order of value exceeds the  noise level.

In Fig.~\ref{7}b we show the enlarged part of charge oscillations in the time interval (186900, 187000). We can
see clear here that the charge amplitude in the second layer is smaller than in the first one. It shows that the
LPW is not the $\pi$-mode, where $k=\pi/d$ and the amplitude is the same in all layers. As we mentioned above,
the LPW with $k=8\pi/9d$ is created in the stack with 9 coupled JJ at $\alpha=1$ and $\beta=0.2$. In
Fig.~\ref{7}c we demonstrate the results of FFT analysis for charge oscillations in the first S-layer at
$I=0.5714$. The LPW frequency is $\omega_{LPW}/2\pi \omega_p = 0.22583$ which  is practically twice smaller than
the Josephson frequency $\omega_J/2\pi \omega_p=0.45166$.

\subsection{Growing region}
\begin{figure}[ht!]
 \centering
\includegraphics[height=30mm]{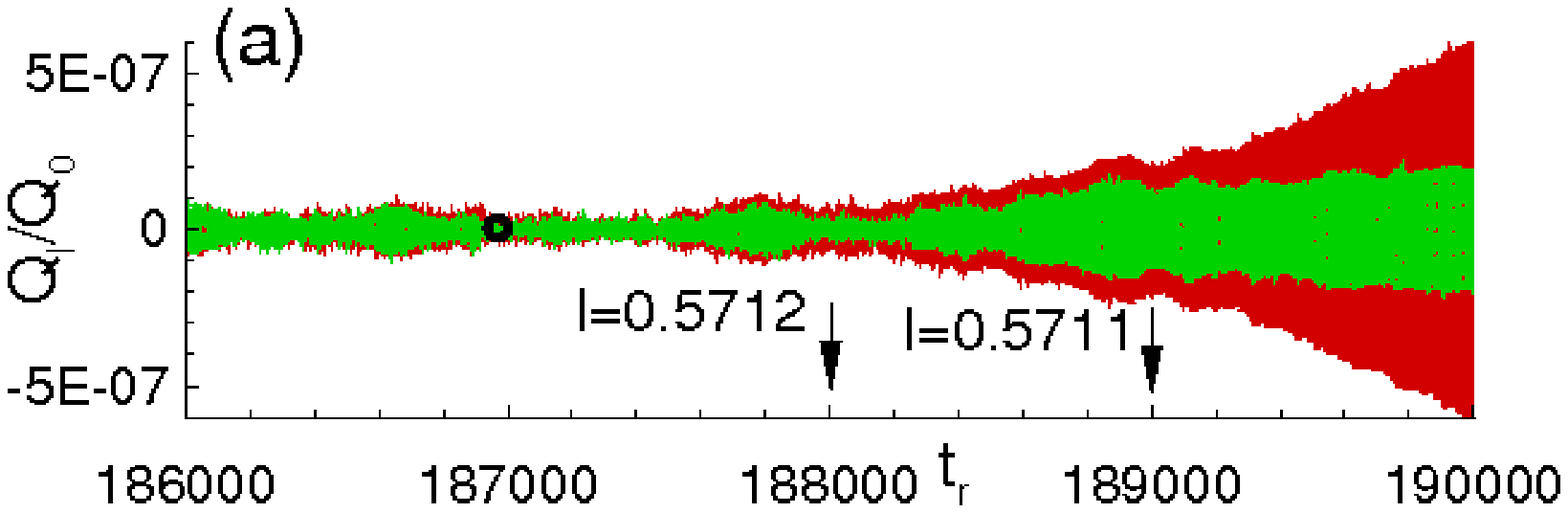}
\includegraphics[height=30mm]{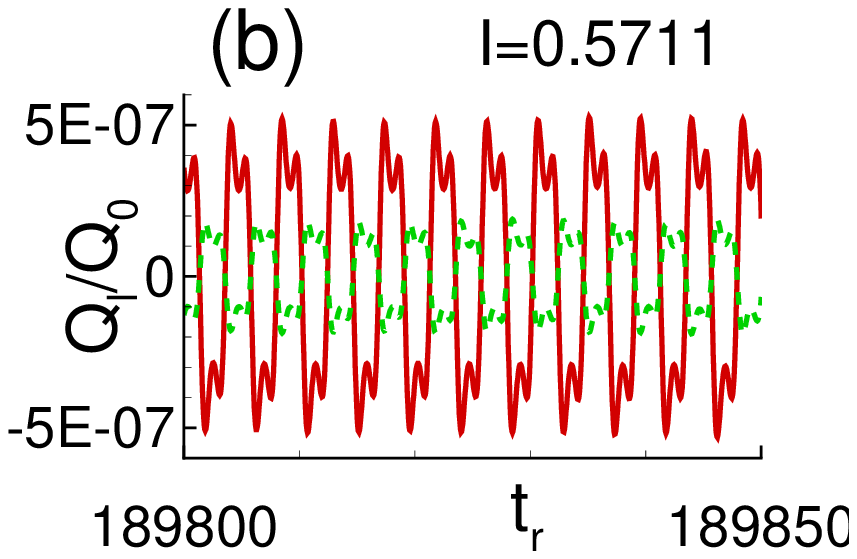}\includegraphics[height=30mm]{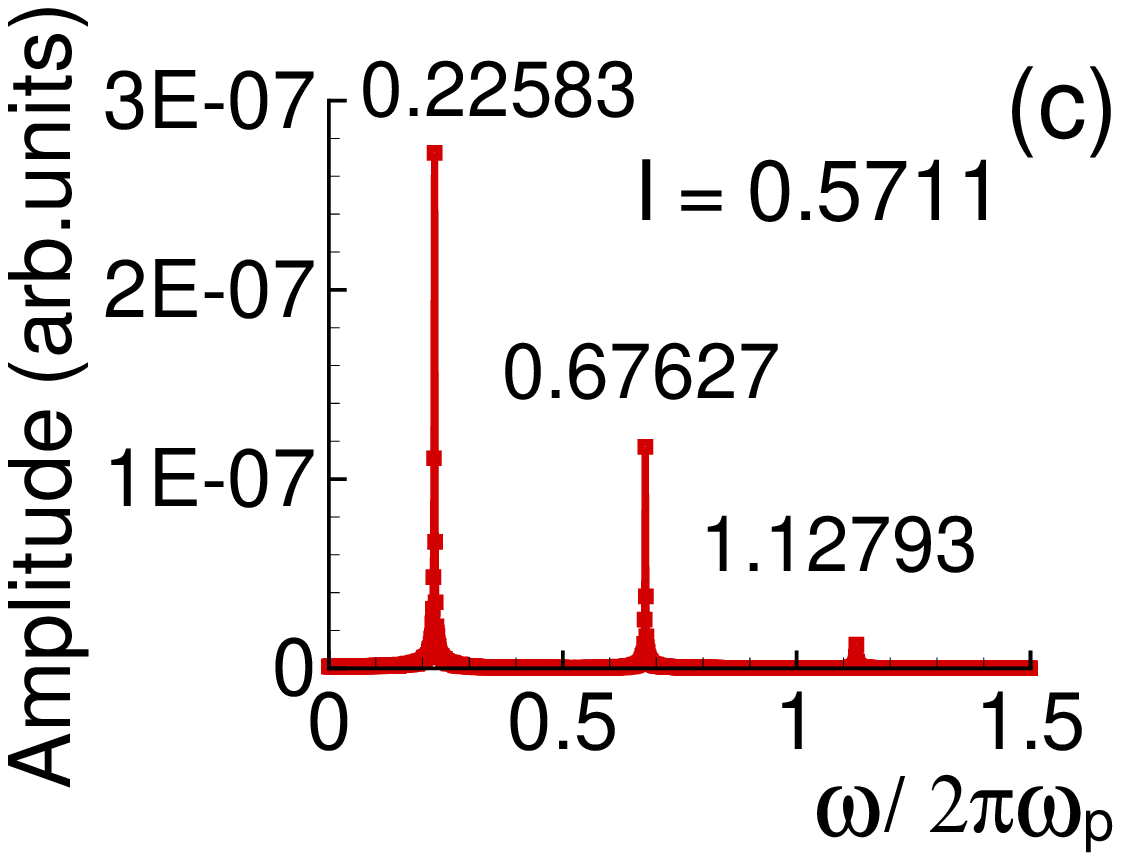}
\caption{(Color online) (a) The characteristic parts of charge-time dependence in the stack with 9 IJJ in
growing region; (b) The enlarged part of (a) in time interval 189800-189850. The thick line shows the charge in
the first S-layer of the stack, the thin line corresponds to the charge in the second S-layer; (c ) Results of
FFT analysis of the charge oscillations  at bias current  $I=0.5711$. }\label{8}
\end{figure}

In Fig.~\ref{8}a and Fig.~\ref{8}b we demonstrate the charge oscillations and some enlarged part for the
beginning of the region with growing amplitude. Fig.~\ref{8}b demonstrates clear the fact that  the amplitudes
of the charge oscillations in the neighboring layers in the growing region is essentially different.
Fig.~\ref{8}c shows the results of FFT analysis of the charge oscillations in the first S-layer at bias current
$I=0.5711$. The peaks are very narrow and  sharp as it should be at the  resonance condition. In this region the
Josephson frequency is equal to the LPW frequency very precisely.

So, we may distinguish clearly  four different stages in time development of LPW: the fluctuation region (LPW in
short time interval on the noise level); the island  region (with charge oscillations exceeding the noise
level); the alternating amplitude region (all islands are joined), and the growing region (charge oscillations
with growing amplitude).

\section{Effects of noise level and external radiation on the LPW nucleation}
\begin{figure}[ht!]
 \centering
\includegraphics[height=70mm]{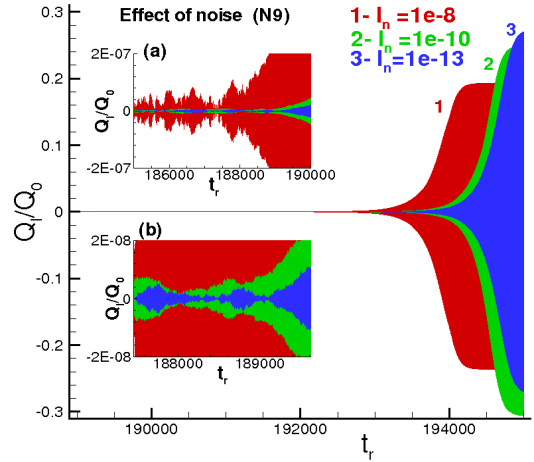}
\caption{(Color online)  Time dependence of the charge oscillations in the first S-layer of  stack with nine IJJ
at different values of the  noise amplitude in  bias current. The insets (a) and (b) show the enlarged beginning
parts of the growing region. }\label{9}
\end{figure}
\begin{figure}[ht!]
\centering
\includegraphics[height=60mm]{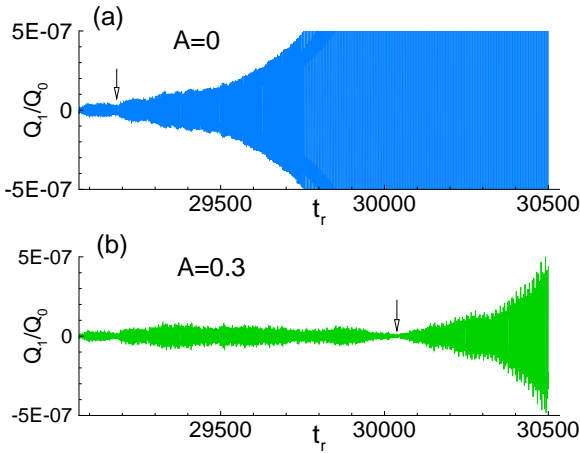}
\caption{(Color online)  (a) Effect of radiation on the  LPW nucleation in the stack with 9 IJJ at $A=0$; (b)
The same at $A=0.3$. } \label{10}
\end{figure}

Let us  briefly discuss the possibility  of affecting the process of LPW's nucleation. First we present the
results concerning the influence of the level of noise in bias current. As it was mentioned above, we add a
small noise with amplitude $10^{-8}$ in bias current in our calculations. The noise stimulates the appearance of
difference of the phase differences in neighbor junctions. It leads to some distribution of the charge in the
S-layers along the stack. At current values much larger than $I_{bp}$  this distribution of charge attenuates in
time. Approaching the breakpoint, the corresponding oscillations are increasing, especially  in the growing
region.

Fig.~\ref{9} demonstrates the time dependence of the charge in the first superconducting layer of the stack with
nine coupled JJ in the beginning of the growing region for three values of the noise amplitude: $10^{-8},
10^{-10}$ and $10^{-13}$. It shows a remarkable fact that the noise with larger amplitude creates the LPW at
smaller time (current) values, than the noise with the smaller amplitude. The inset (a) allows us to see clear
that the growing region starts first in case of the noise amplitude equal to $10^{-8}$ and the inset (b)
demonstrates that for the noise amplitude $10^{-13}$ the growing region starts in the last turn. This result is
understandable: it's  easier to arrange a difference between neighbor junctions by larger value of the noise
amplitude.

The another possibility to affect the nucleation process is the effect of the external radiation. If we
irradiate a stack of Josephson junctions with microwave of frequency $\omega_r$, it causes an alternating
current through the stack. This alternating current can be described as the additional applied current $A
\sin(\omega_r t)$, where A is a dimensionless amplitude.

In Fig.~\ref{10} we demonstrate the effect of radiation on the charge oscillations in the first S-layer in the
beginning of the growing region. We see that the radiation with frequency $\omega/\omega_p=0.5$ and amplitude
$A=0.3$ (Fig.~\ref{10}b) shifts the starting point of growing region (shown by arrow) in comparison with case
$A=0$ (Fig.~\ref{10}a). Our results show that the change in the frequency leads to the shift of this point and
additional resonance features. This effect of radiation will be considered in detail somewhere else.

We would like to mention here that the nucleation process can be affected by change in temperature as well. The
change in the McCumber parameter should lead to the creation of the LPW with another wave number\cite{sm-prl07}
and, correspondingly, to the nucleation process specific  to this wave number .

In summary, we studied the nucleation of the longitudinal plasma wave in the coupled system of IJJ in HTSC and
found the different stages in its process of development. The answer to the question concerning the
correspondence between the breakpoint's position in CVC and the parametric resonance region in time dependence
of the charge on the S-layers is found. We showed that the position of the breakpoint on the outermost branch of
CVC is related to the region with sharp increase of the amplitude of charge oscillation in the superconducting
layers. We demonstrated that the onset of the growing region can be shifted by noise in the bias current and
microwave radiation. These effects open a way to regulate the process of LPW nucleation in the stack of IJJ.

We thank M. R. Kolahchi, R. Kleiner, M. Suzuki and S. Flach for fruitful discussions. This research was
supported by the Russian Foundation for Basic Research, grant 08-02-00520-a.

\end{document}